\documentclass[twocolumn]{aastex6}

\usepackage{graphicx}
\usepackage{amsmath}
\newcommand{\flbox}{\ensuremath{F_\mathrm{box}}}
\newcommand{\fsky}{\ensuremath{F_\mathrm{sky}}}
\newcommand{\mlim}{\ensuremath{m_{R}^{\mathrm{lim}}}}
\newcommand{\mzp}{\ensuremath{m_{R}^{\mathrm{zp}}}}
\newcommand{\phiiq}{\ensuremath{\Phi_\mathrm{IQ}}}
\newcommand{\frat}{\ensuremath{\vartheta_\mathrm{ratio}}}

\shorttitle{The Palomar Transient Factory survey efficiency}
\shortauthors{Frohmaier et al.}

\begin{document}

\title{Real-Time Recovery Efficiencies and Performance of the Palomar Transient Factory's Transient Discovery Pipeline}

\author{C. Frohmaier, M. Sullivan}
\affil{Department of Physics and Astronomy, University of Southampton, Southampton, Hampshire, SO17 1BJ, UK}
\email{c.frohmaier@soton.ac.uk}
\and
\author{P. E. Nugent, D.A. Goldstein}
\affil{Department of Astronomy, University of California, Berkeley, CA, 94720-3411, USA; Lawrence Berkeley National Laboratory, Berkeley, CA, 94720, USA}
\author{J. DeRose}
\affil{Kavli Institute for Particle Astrophysics \& Cosmology, P. O. Box 2450, Stanford University, Stanford, CA 94305, USA; Department of Physics, Stanford University, 382 Via Pueblo Mall, Stanford, CA 94305, USA}

\begin{abstract}
  We present the transient source detection efficiencies of the
  Palomar Transient Factory (PTF), parameterizing the number of
  transients that PTF found, versus the number of similar transients
  that occurred over the same period in the survey search area but that
  were missed. PTF was an optical sky survey carried out with the
  Palomar 48-inch telescope over 2009--2012, observing more than 8000
  square degrees of sky with cadences of between 1 and 5 days,
  locating around 50,000 non-moving transient sources, and
  spectroscopically confirming around 1900 supernovae. We assess the
  effectiveness with which PTF detected transient sources, by
  inserting $\simeq$7 million artificial point sources into real PTF
  data. We then study the efficiency with which the PTF real-time
  pipeline recovered these sources as a function of the source
  magnitude, host galaxy surface brightness, and various observing
  conditions (using proxies for seeing, sky brightness, and
  transparency). The product of this study is a multi-dimensional
  recovery efficiency grid appropriate for the range of observing
  conditions that PTF experienced, and that can then be used for
  studies of the rates, environments, and luminosity functions of
  different transient types using detailed Monte Carlo simulations. We
  illustrate the technique using the observationally well-understood
  class of type Ia supernovae.
\end{abstract}


\section{Introduction}
\label{sec:introduction}

The last decade has seen a revolution in the study of the optical sky
in the time domain. Several large-area `rolling searches' -- for
example, Pan-STARRS 1 \citep[][]{2010SPIE.7733E..0EK}, the Catalina
Real-Time Transient Survey \citep[][]{2009ApJ...696..870D}, the La
Silla Quest Variability Survey \citep[][]{2013PASP..125..683B}, and
the Palomar Transient Factory \citep[PTF\footnote{\url{http://www.ptf.caltech.edu/}};][]{2009PASP..121.1334R} -- have
repeatedly surveyed the sky on time-scales from minutes to hours, days
and years.  These surveys, together with dedicated spectroscopic
follow-up programs \citep[e.g.,][]{2015A&A...579A..40S}, have
discovered thousands of galactic and extra-galactic astrophysical
transients each year, filling in new and previously unexplored regions
of the time-domain phase space.

Understanding the efficiency with which these surveys operate and
detect objects is of paramount importance in understanding the astrophysics of the transient populations that they uncover.
For every transient that is detected, it is important to know how many
events with the same properties were not detected during the survey
period. There are many reasons why transients can be missed or not detected by surveys, beyond
simple Malmquist bias effects. For example, the observational cadence of the survey
may be too long to detect rapidly evolving events; gaps in observing
as a result of poor weather, seeing, or technical problems may occur;
some parts of the survey area may be inaccessible due to saturated
foreground stars, gaps between CCDs, or bad pixels; the detection sensitivity may change
as a function of the lunar cycle or other variables; inefficiencies in
the complex data reduction and transient detection pipelines may result in
transients of any brightness being lost. All surveys will therefore
make an inevitably incomplete sampling of the transient population,
which will consequently impact the determination of transient
volumetric rates, luminosity functions, the dependence of the
transient on the underlying stellar populations, and, in the case of
cosmological studies using supernovae, the measured cosmological
parameters.

These effects and losses can be corrected for, if the efficiency of a
survey can be determined. Studies that attempt this require
large-scale simulations that can be computationally very expensive.
They invariably work via the insertion of `fake' transients into a
survey imaging data stream, passing the adjusted data through the same
survey detection pipeline as used to find real transients, and
assessing the degree to which the fake transients can then be recovered. This
can be done either `offline' once a survey has been completed
\citep[e.g.,][]{2002ApJ...577..120P,2010AJ....140..518P}, or in
real-time while the survey is operational and the data being collected
\citep[e.g.,][]{2008AJ....135..348S,2015AJ....150..172K}.  The fake
events are usually designed to replicate the properties of the entire range of
transients that might be detected, from their apparent magnitude to
their host galaxy environment and local surface brightness.

In this paper, we present the survey and detection efficiencies for
the real-time difference imaging pipeline of the Palomar Transient
Factory
\citep[PTF;][]{2009PASP..121.1334R,2009PASP..121.1395L},
with a particular view to the study of supernovae and supernova-like
transients.  PTF is an automated optical sky survey operating at the
Samuel Oschin 48-inch telescope (P48) at the Palomar Observatory, and
is specifically designed for transient detection. The initial phase of
PTF, on which this paper is based, conducted an optical sky survey
over 8000\,deg$^2$ from 2009--2012 operating with cadences
designed to span one to five days. The survey located nearly 50000
non-moving astrophysical transients, and spectroscopically confirmed
1900 supernovae over this period, leading to large samples of supernovae of
different types
\citep[e.g.,][]{2014MNRAS.444.3258M,2015ApJ...799...52W,2016ApJ...820...33R}.

Determining the efficiency of PTF in order to fully exploit these
samples for population studies is challenging. Surveys focused on the
detection and study of high-redshift type Ia supernovae (SNe Ia),
e.g., the Dark Energy Survey \citep{2015AJ....150..172K} and the
Supernova Legacy Survey \citep{2010AJ....140..518P}, often use a Monte
Carlo approach to determining detection efficiencies, synthesizing the
light curves of thousands of supernovae over a particular observing
season, and inserting fake point sources into each image with the
correct photometric properties following the evolution of the
synthesized events. This allows the simultaneous determination of both
the efficiency on any given epoch, and the recovery efficiency of the
underlying SN Ia population.  While this is practical for surveys that
observe a limited number of fixed fields with a primary interest in
one particular supernova type, it does not translate effectively into
a survey such as PTF, where we wish to study the populations of any
supernova-like transient that PTF could detect.

Indeed, PTF presents its own unique challenges. PTF covered a large
area of sky (approximately 8000 deg$^2$ in the 3-5 day cadence experiment), operated 9 months per
year for four years, and was allocated around 80\% of the P48 time
over this period, achieving an observing efficiency of $>$50\%
open-shutter in good conditions \citep{2009PASP..121.1395L}.  During
this period, $\geq2.2\times10^{6}$ images were taken and processed generating just over 1PB of total data in the pipeline including reference, subtraction and noise images, as well as a nearly 1TB database storing the metadata from every image and all candidate transient detections. It is thus impractical to insert fakes into all of these images in sufficient
numbers to study the recovery efficiency on a per-field basis.

Our approach to determining supernova rates and population statistics
in PTF is therefore a two-step process. In the first step, detailed in
this paper, we choose a single representative field in PTF observed
hundreds of times over the four years, with observing conditions that
sample the full range that PTF experienced. We insert millions of fake
point sources (`fakes' or `fake SNe') into every image of this single area, pass them through the
detection pipeline, and construct a recovery efficiency grid as a
function of variables such as the transient brightness, image
photometric zeropoint, and seeing.

The second step then uses this grid together with Monte Carlo
simulations of particular transient types in the PTF survey. In these
simulations, fakes are not inserted into images, and instead the PTF pipeline database, which contains the observing conditions of every PTF image, reference and subtraction, is queried together with the detection efficiency grid described above. The recovery efficiency for any event can then be calculated from interpolating the detection efficiency grid at the position corresponding to the transient brightness and the observing conditions taken from the PTF database.
This method achieves a computational saving over the traditional approach of inserting transient-specific fakes into every image. The slowest element of the analysis is the image manipulation and source detection of the fakes. An advantage of our technique is that this only needs to be performed once, regardless of the different transients we want to study. We outline this procedure in this paper, but describe the specific application to particular SN types in later articles.

A plan of the paper follows. In section \ref{sec:det_eff} we describe
the sample of PTF data on which we conduct our fake transient
experiments, and show that these data are representative of the entire
survey. We describe the method with which fake point sources are
added into the observational data, and the process of recovering the fake sources using the PTF
real-time detection pipeline, showing that the fakes are reliable
probes of the survey detection efficiency. The recovery fractions are
quantified in section \ref{sec:RF} as both single and multi-dimensional
functions of the observing parameters and of the fake properties
themselves.  Finally, in section \ref{sec:survey_ops} we demonstrate
our method of simulating the survey as a time-dependent sky
probability map of detections with a demonstration using a real
astrophysical transient population, SNe Ia. Throughout, where relevant we assume a flat $\Lambda$CDM Universe with $\Omega_\mathrm{M}=0.3$ and a Hubble constant $H_0$=70\,km\,s$^{-1}$\,Mpc$^{-1}$, and work in the AB photometric system \citep[e.g.,][]{1983ApJ...266..713O}.

\section{Recovery efficiencies in PTF}
\label{sec:det_eff}

In this section we detail the pipeline that PTF uses to find new
transient objects in its imaging data, and describe our method of
testing the performance of this pipeline (the `recovery efficiency').
PTF, like many other sky surveys, finds astronomical transients
through a process of image subtraction. In this process, a new
`science' image taken on a given night is astrometrically and
photometrically aligned to a `reference' template image constructed
from an average of several images taken previously in good conditions.
The point-spread function (PSF) of the two images is then matched, and
the reference image subtracted from the new science image. This leaves
an image containing only astrophysical transients that have changed in
brightness or position between the two images, as well as subtraction
artefacts due to imperfections in the image subtraction process, and
other artefacts such as cosmic-rays. Different astrophysical transients can be
characterized by a different spatial and temporal evolution: as a
trivial example, asteroids move quickly across a field, whereas
supernovae are static but change in brightness.  These differences
allow for machine classification to select and reject candidate
objects found in the image subtractions.  We describe each of these
steps in turn.

\subsection{The PTF transient detection pipeline}
\label{sec:sub_pipe}

The PTF detector is the CFH12k instrument mounted at the Samuel Oschin
48-inch telescope (the P48) at the Palomar Observatory. The CFH12k was
previously mounted on the Canada--France--Hawaii Telescope, and has 11
functional 2048x4096 pixel CCDs\footnote{The 12th CCD, CCD03, failed
  early in the PTF program and was not replaced.} arranged in two rows
of six, giving an active field of view of 7.3\,deg$^{2}$ during the
PTF survey, with a pixel scale of 1.01\arcsec\ pixel$^{-1}$.  First
light occurred on 2008 December 13, with the survey commencing on 2009
March 1 and continuing until 2012 December 31. The 3--5 day cadence
experiment, which forms the primary dataset for our study, ran from 1
March until 31 October each year, using around 65\% of the available
P48 telescope time. PTF operated primarily using a Mould $R$ filter
($R_{\mathrm{P48}}$) and a Sloan Digital Sky Survey (SDSS)
$g^{\prime}$ filter ($g_{\mathrm{P48}}$) with 60\,s exposure times.
The majority (83\%) of the data were taken with the $R_{\mathrm{P48}}$
filter, and we consider only these data in this study.

The PTF real-time transient detection pipeline is hosted at the
National Energy Research Scientific Computing Center (NERSC). A
description of the pipeline can be found in
\citet{Nugent2015,2016PASP..128k4502C}, and a brief overview is given
here. The pipeline performs bias-subtraction and flat-fielding, and
determines approximate astrometric solutions through
astrometry.net\footnote{\url{http://astrometry.net}}. The
\textsc{sextractor} object detection program
\citep{1996A&AS..117..393B} detects and measures the fluxes of objects
in each image, and compares to the United States Naval Observatory
(USNO)-B1 catalogs \citep{2003AJ....125..984M} to calculate the
photometric zero-point.

A significant amount of additional metadata are generated by the
real-time pipeline describing the context and properties of each
CCD image (characterized by over 90 variables), and we make extended use of these image metadata in this paper. In
particular, these data describes the effect of the observing conditions on the images. The metadata, stored for every CCD, include:

\begin{enumerate}

\item The 3\,$\sigma$ limiting apparent magnitude on each unsubtracted image in the
  $R_{\mathrm{P48}}$ filter (\mlim),

\item The zeropoint to calibrate instrumental magnitudes to the USNO-B1 photometric system (\mzp),

\item The Full Width at Half Maximum (FWHM) of the image PSF (hereafter referred to
as the image quality, IQ). Additionally, the ratio of the IQ in the science image to the IQ of the reference image \phiiq\ is stored,

\item The median sky level in counts (\fsky),

\item The airmass of the observations,

\item The mean ellipticity of sources in the image,

\item The moon illumination fraction, with 0 denoting new moon, and -1 or 1 denoting full moon.

\end{enumerate}

Following this basic data reduction, the pipeline performs the image
subtraction. At regular intervals during the survey operations, the
reference images were created and updated from previous observations
of each field. The new image and the corresponding reference image are
astrometrically aligned using \textsc{scamp}
\citep{2006ASPC..351..112B} and the reference image resampled to the
same pixel system as the new image using \textsc{swarp}
\citep{2002ASPC..281..228B}. The subtraction package
\textsc{hotpants}\footnote{\url{http://www.astro.washington.edu/users/becker/v2.0/hotpants.html}}
is then used to create a subtraction image from the new and reference
images. Object detection on this subtraction image is performed using
\textsc{sextractor}, and the output fed into the machine learning
algorithm of \citet{2012PASP..124.1175B} to assign an Real-Bogus (RB)
score to all the detections.

The machine learning is necessary for the automated discovery and
classification of transient objects due the the vast number of
pseudo-candidates extracted in the subtraction images. Only 0.1\% of
the candidates in any given subtraction would be considered to have an
astrophysical origin, and this, coupled with the 1-1.5 million
candidates stored in the PTF database each night, presents an
overwhelmingly large challenge for human scanners to review
everything. The machine learning algorithms developed for PTF are
designed to make a statistically supported assertion as to whether a
candidate is astrophysically real or `bogus'. The algorithm was
trained on the assessments of human scanners who operated during
commissioning and early operations of PTF. These scanners were asked
to assess cut-out images of candidates from image subtractions, and to
assign a score to that candidate from 0 (bogus) to 1 (real). From
this, a set of `features' were determined from the \textsc{sextractor}
output catalogs which could be used to assign an RB score to a
candidate so that it best replicates the results of the human
scanners. A full list of the features can be found from Table 1 in
\citet{2012PASP..124.1175B}.

\subsection{Simulations}
\label{sec:simulations}

Our simulations are designed to test the performance of the real-time
PTF pipeline described above, therefore the data products we generate from this study\footnote{The catalog of fakes used to generate the efficiency grids in Section \ref{sec:RF} are available in a persistent directory
    \dataset[10.5258/SOTON/D0030]{http://doi.org/10.5258/SOTON/D0030}.} must be used only with the real-time outputs. Any additional image
calibration, external to the real-time pipeline, would change the results we find for the transient detection pipeline. For a given set of transient properties and observing conditions, the `recovery efficiency' $\epsilon$ is defined as the ratio of the number of transients found by a survey, to the total number of similar transients that occurred within a fixed sky area. 
That is, it is the probability that an astrophysical event with a
given set of properties is recovered on a given epoch. We refer to
this as the `single epoch' recovery efficiency, and it is a complex
multi-dimensional function of transient properties (e.g., the transient
apparent magnitude $m_R$), astrophysical environmental properties (e.g., local host galaxy surface brightness), and observing conditions (e.g., IQ, $m_{R}^{\mathrm{lim}}$, etc.). Although some
surveys monitor such a recovery efficiency in near real-time by
inserting artificial point sources into the data as it is taken each
night \citep[e.g., the DES SN program;][]{2015AJ....150..172K}, this
approach was not used in PTF due to the heavy computational demand of
doing this on a near-continuous data stream.
 
Our analysis was performed on PTF data taken between 2009 and 2012
when the survey was fully operational. We evaluate the recovery
efficiency by inserting a population of artificial point sources (`fakes')
into the PTF imaging data. The resultant images are then
treated identically to a new observation, and processed through the
same transient detection pipeline as used during the survey
(Section~\ref{sec:sub_pipe}), including the machine learning
classification. A comparison between the input fake population and the
population recovered by the pipeline then provides information on the
recovery efficiency on any epoch as a multi-dimensional function of the
fake's properties and observing parameters that describe the data.

The computational load of this process -- inserting fakes and running the detection pipeline on
the resulting image -- is high, taking around 7.7\,s per PTF exposure (running
the 11 CCDs of each exposure in parallel). Thus to analyze every image used by PTF in the image subtraction pipeline once, would require $>$150 days of supercomputer time. In reality, many additional 
iterations on each image would be required in order to accumulate the
necessary statistics on each epoch, further increasing the required computing time.

Instead, we choose to perform our analysis on a single PTF field, but
one that sampled a representative range of observing conditions
experienced by the survey. We choose PTF field 100019, observed 1290
times over the survey duration.  This field contains the galaxy M101
that hosted the SN Ia SN\,2011fe\footnote{Although the typical
  exposure time in PTF is 60\,s, due to the brightness of SN\,2011fe
  (reaching $m_R\sim10$\,mag), the exposure time for observations of
  field 100019 were shortened during the period that SN\,2011fe was
  bright, to avoid saturation of the SN.  These shorter exposures,
  which make up 15\% of the field 100019 observations, are discarded
  from our analysis as they are not representative of PTF as a
  whole.} \citep{2011Natur.480..344N}, and was
observed with an almost daily cadence as part of the `dynamic cadence'
PTF program \citep{2009PASP..121.1395L} in order to study novae and
`fast and faint' transients
\citep[e.g.,][]{2012PASA...29..482K}.

\begin{figure}
\epsscale{1.0}
\centering
\includegraphics[width=\linewidth]{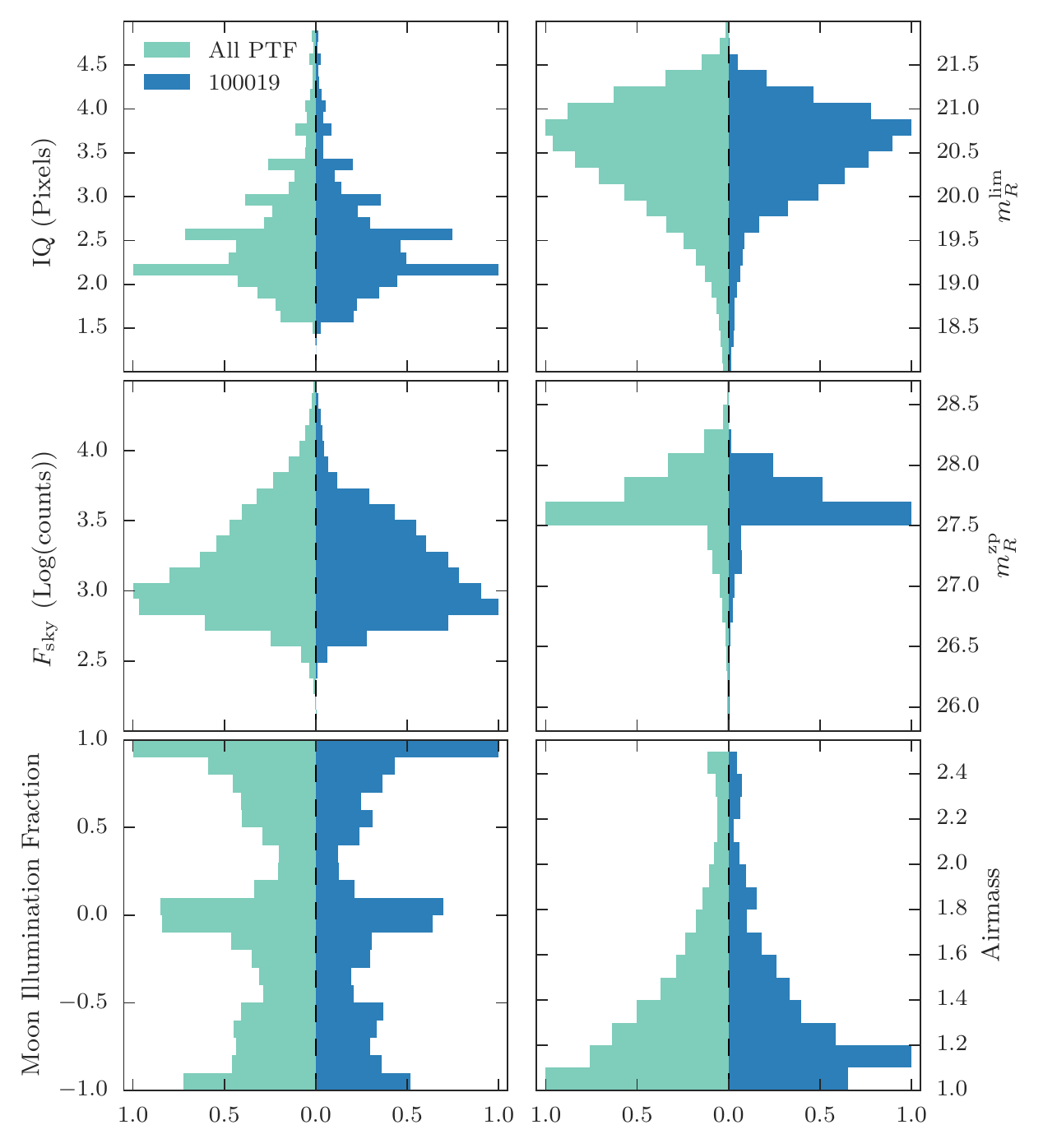}
\caption{The (renormalized) distributions of the image metadata and
  observing conditions (Section~\ref{sec:sub_pipe}) across the entire
  PTF survey (left, light-shaded histograms) compared to those of the PTF field
  100019 used in our recovery efficiency simulations (right, dark-shaded histograms). The top left panel shows the comparison for the image
  quality (IQ), the top right panel the limiting magnitude
  (\mlim), the center left panel the median sky
  counts (\fsky), the center right panel the photometric zeropoint (\mzp), the lower left panel the moon illumination fraction (0 = new moon, -1,1 = full moon), and the lower right the airmass of the observation.}
\label{fig:conditions}
\end{figure}

Figure \ref{fig:conditions} shows how the image metadata and observing
conditions of field 100019 compare to that experienced by the PTF
survey as whole. While identical distributions are not required, it is important that the full range of conditions is sampled by field 100019, and that the distributions are similar, so that the computational resources are used efficiently. It is clear in Figure \ref{fig:conditions} that there is a good agreement between our chosen field and that of PTF as a whole.

\subsubsection{Selecting point sources}
\label{sec:fake-point-sources}

Our fakes are sampled from real point-sources located in each image.
We use \textsc{sextractor} to locate the 20 brightest, unsaturated,
and isolated point sources (i.e., `stars'), ensuring each is
$>50$ pixels from the CCD edge. Our selection is based on
the \textsc{sextractor} neural network \textsc{class\_star}
classifier, which assigns every object a value from 0 (not star-like)
to 1 (star-like). This cut removes galaxies and cosmic rays from our
fakes catalog, which we confirmed by visual inspection from a random sample of 1084
candidate stars. We do note, however, that a small fraction of the visually inspected stars show some `blooming' into adjacent pixels. This contamination is difficult to filter out as these stars still receive a high \textsc{class\_star} value in \textsc{sextractor}. For our selected stars sources, 99.5\% of the objects have a \textsc{class\_star} score \textgreater0.92.

Our fakes are then constructed by `clone-stamping' these bright
stars: we take a box of 9 pixels on a side that encloses the PSF,
subtract the local \textsc{sextractor} background, and re-scale the
star to the desired fake apparent magnitude ($m_R$). This method
ensures that the fakes have a PSF that is both representative of
real objects in the image, but also carries the intrinsic variation of
the PSF (the object-to-object variation) within the simulation.  We
generate fakes with a uniform magnitude distribution from
$m_R$=15--22\,mag. We additionally enforce the condition that each
fake must be a least one magnitude fainter than the original star from
which it was generated.

\subsubsection{Inserting fakes into the data}
\label{sec:inserting-fakes-into}

A key consideration when inserting the fakes into the PTF data is
that the presence of these `extra' sources does not distort the
machine learning classification process.  One of the 28 metrics
\citep{2012PASP..124.1175B} that goes into the RB score is the spatial
density of good candidates, defined as the `ratio of the number of
candidates in that subtraction to the total usable area on that
array'. Thus, saturating an image with an artificially high density of
fakes may lead to unrepresentative RB scores.  A secondary effect
is that adding too many fakes into an image could affect the
astrometric alignment of the science image to the reference, and thus
cause an increased number of subtraction artefacts.

\begin{figure}
\epsscale{1.0}
\centering
\includegraphics{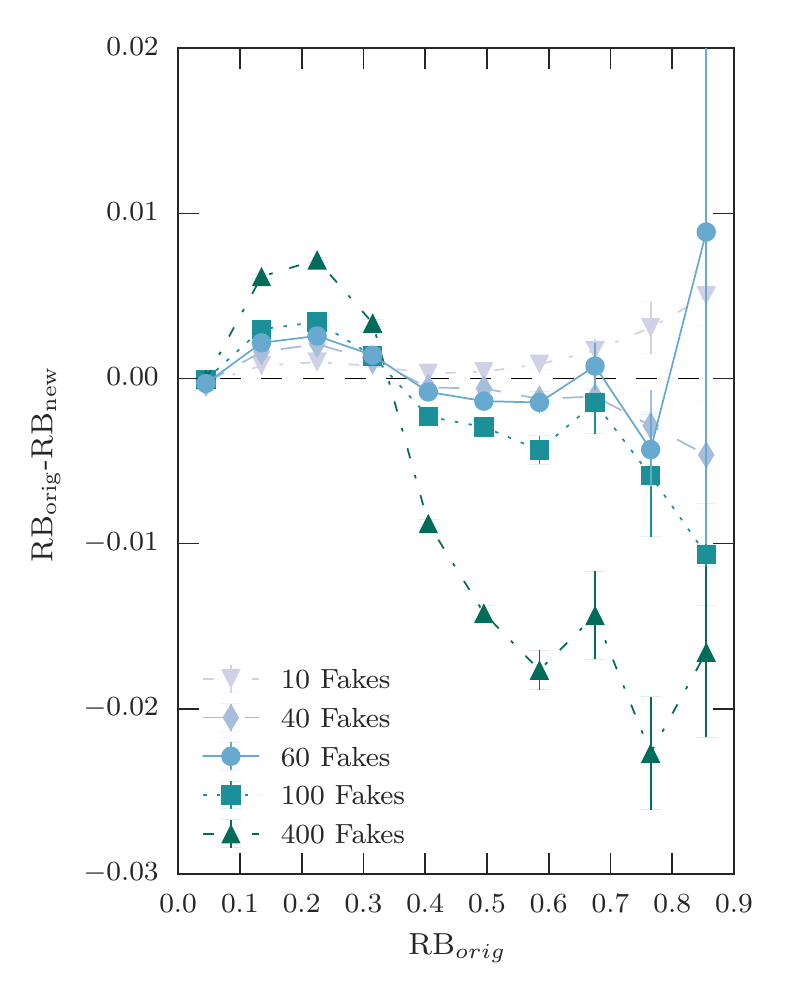}
\caption{The original Real-Bogus (RB; Section~\ref{sec:sub_pipe}) scores (RB$_\mathrm{orig}$) of the real
  candidate objects in our sample images, compared to the RB scores of
  the same objects with different numbers of fakes added to the same images
  (RB$_\mathrm{new}$). The figure shows cases where 10, 40, 60, 100
  and 400 fakes have been added to the images.}
\label{fig:ML_Diff}
\end{figure}

We therefore investigated, using a random sample of 281 images made available to us for pipeline development from the tape archives,  how the
addition of fakes changed the RB scores of real candidates in the
images.  In Figure~\ref{fig:ML_Diff}, we compare our baseline RB
scores of real candidates (when there are no fakes in an image) with
the RB scores of the same candidates but with an increasing number of
fakes added.  We find that even a small number of fake objects
slightly distorts the RB scores; however these effects remains
negligible when of order tens of fakes are added, only becoming
important with $>$100 fakes.  Based on our analysis, we consider 60
fake objects per image to be a satisfactory compromise between
maximizing our computational efficiency and distorting the RB scores.
We also note that even with 400 fakes per image, the astrometric
alignment to the reference image was not changed.

\subsubsection{Fake Point Source locations}
\label{sec:fake-sn-locations}

Most real astrophysical transient events occur within an associated
host galaxy. However, if our fakes were added to random locations
on the sky, then the majority would instead be placed in host-less
regions, and consequently would provide poor statistics on the
recovery efficiency as a function of host galaxy parameters, such as
local surface brightness. This would require us to perform many more
fake point-source simulations in order to adequately map this parameter space.

We therefore choose to bias the locations of our fakes to ensure
that 90\% of them are placed within a detected galaxy. To select a
host for these fake point sources, the \textsc{sextractor} catalogs were used
to randomly choose galaxies in each image, with the galaxy pixel
positions given by ($x_\mathrm{gal}$, $y_\mathrm{gal}$). A fake is
then added at a pixel position ($x_\mathrm{SN}$, $y_\mathrm{SN}$) at
an elliptical radius $R$ within the isophotal limit of each galaxy.
The elliptical shape parameters are measured by \textsc{sextractor},
defined by the semi-major ($r_A$) axis, the semi-minor ($r_B$) axis, and
the position angle ($\theta$), with $R$ given by
\begin{multline}
\label{eqn:gal_ell}
R^2=C_{xx}(x_\mathrm{SN} - x_\mathrm{gal})^2 + C_{yy}(y_\mathrm{SN} - y_\mathrm{gal})^2+
\\
C_{xy}(x_\mathrm{SN} - x_\mathrm{gal})(y_\mathrm{SN} - y_\mathrm{gal})
\end{multline}
where $C_{xx}=\cos^2(\theta)/r_A^2+\sin^2(\theta)/r_B^2$,
$C_{yy}=\sin^2(\theta)/r_A^2+\cos^2(\theta)/r_B^2$, and
$C_{xy}=2\cos(\theta)\sin(\theta)(1/r^2_A-1/r_B^2)$. A value of
$R\sim3$ corresponds to the isophotal limit of each object. The
location of each fake is not refined further, for example to follow
a galaxy surface brightness profile. The remaining 10\% of the fakes
were added into blank regions of the sky. We also ensure that a fake
is not within 40 pixels of another fake, regardless of whether it is
in a galaxy or not.

\subsection{Fake supernova recovery}
\label{sec:ob_rec}

The simulation method described above is applied 10 times to all
observations of the PTF field 100019 taken over 2009--2012, generating
a sample of $\approx$7$\times10^{6}$ fakes in the data. The product of our simulations are two PostgreSQL\footnote{\url{https://www.postgresql.org/}} database tables. The first stores a complete description of the parameters describing each fake: the spatial location and any host galaxy information, the fake magnitude, and the observing conditions metadata.
The second table stores the output from the real-time detection
pipeline run on the images containing the fakes, including the
machine learning RB scores; i.e., it contains information on which
fakes were recovered by the pipeline (as well as all the real
astrophysical transients and false-positives).

\begin{figure}
\centering
\includegraphics[width=\linewidth]{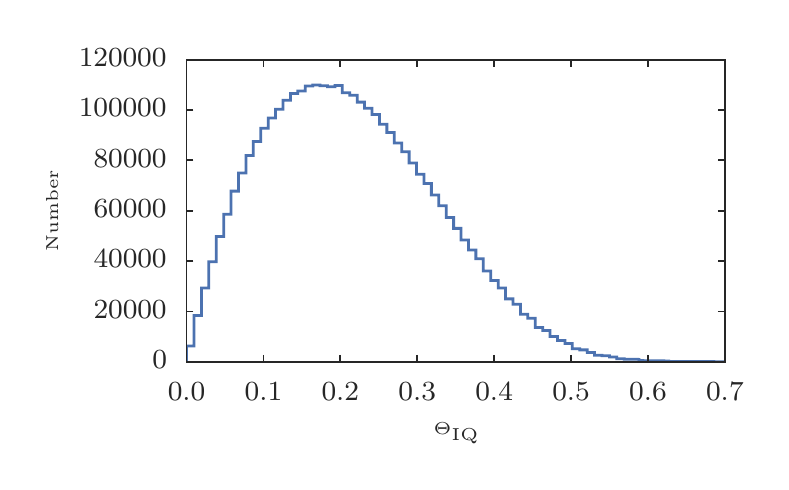}
\caption{The distribution of $\Theta_\mathrm{IQ}$, where $\Theta_\mathrm{IQ}$ is defined as the separation
  between the position of the input fake and the position of the nearest recovered
  candidate, divided by the IQ of the image. We require that
  $\Theta_\mathrm{IQ}<0.6$ in order to consider a fake to be
  recovered by the pipeline.}
\label{fig:Sep}
\end{figure}

To determine whether a fake was recovered by the pipeline, we
perform a spatial matching of the two databases (fake positions versus
recovered positions), and require that any matched fake must have a
RB score $\geq0.07$, the same as during the PTF survey operation
\citep{2012PASP..124.1175B}.  The matching radius between a fake
and a recovered candidate varies with the IQ (seeing), and to remove spurious
associations we define $\Theta_\mathrm{IQ}$ as the ratio of the
separation of a fake and the nearest recovered candidate, to the
IQ.  The histogram of all $\Theta_\mathrm{IQ}$ is shown in Figure
\ref{fig:Sep}, and we enforce $\Theta_\mathrm{IQ}<0.6$ in order to
consider a fake to be recovered. Any fake without a detection
satisfying $\mathrm{RB}\geq0.07$ and $\Theta_\mathrm{IQ}<0.6$ is
considered not recovered.

\subsection{Recovered fake point source properties}
\label{sec:recovered-fake-sn}

\begin{figure}
\epsscale{1.0}
\centering
\includegraphics[width=\linewidth]{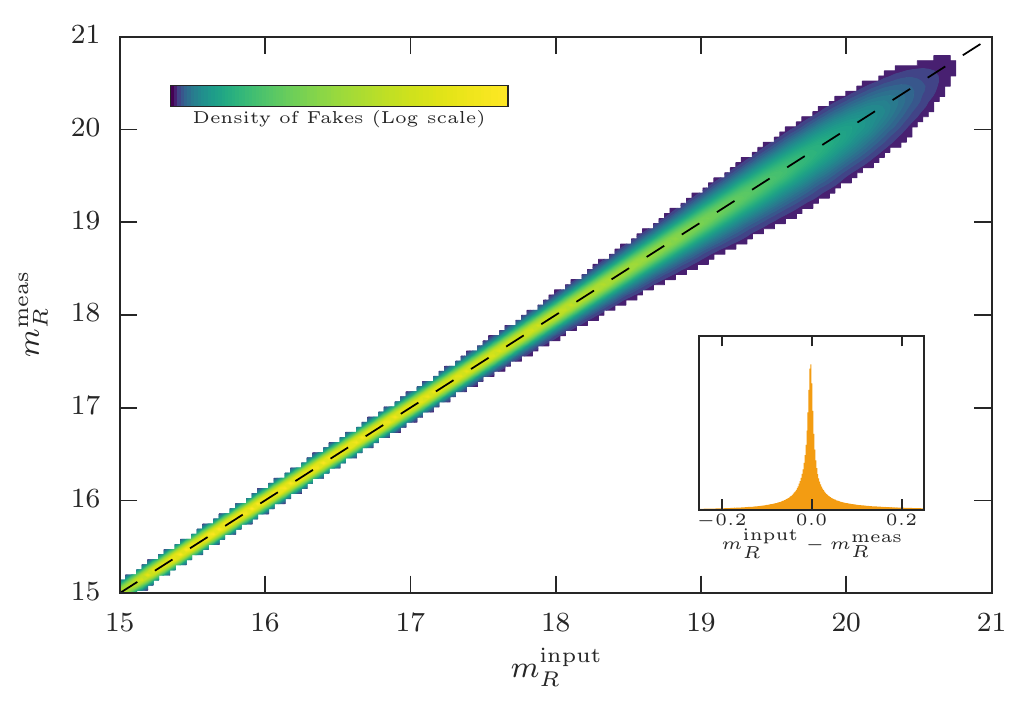}
\caption{The \textsc{sextractor} measured magnitudes
  ($m_R^\mathrm{meas}$) for the recovered fakes compared to their
  input magnitudes ($m_R^\mathrm{input}$). The main panel shows the
  overall comparison with the dashed line denoting a 1:1 agreement,
  and the inset panel shows the distribution of
  $m_R^\mathrm{meas}-m_R^\mathrm{input}$ which is sharply peaked at 0 showing no systematic offset.}
\label{fig:maginout}
\end{figure}

We next compare the recovered fake's magnitude to that input into the
pipeline (Figure \ref{fig:maginout}). Although this is not a critical
part of our analysis, as we do not use the recovered photometry in our
analysis, this test acts as a useful sanity check that our efficiency
pipeline is working as expected, and that the PTF real-time pipeline
itself can recover reasonably accurate photometry. The agreement is
generally good, and as expected, the fainter fake SNe show a larger
scatter between their input and recovered magnitudes as the
signal-to-noise (S/N) decreases; however the overall comparison shows
a good agreement with no systematic offset. We find that 92\% of the
recovered fake magnitudes are within 0.2\,mag of their input
magnitudes, and splitting our fakes into bright objects
($m_R\leq18.5$\,mag) and fainter objects ($m_R>18.5$\,mag) we find
98\% and 77\% of the magnitudes are recovered within 0.2\,mag.  Thus
the PTF real-time search pipeline accurately recovers the input
magnitudes of the fakes.

\section{Recovery statistics}
\label{sec:RF}

We now study the performance of the pipeline in recovering fakes under
different observing conditions, and as a function of the input fake's
properties and location. We use this to motivate the construction of a
multi-dimensional recovery efficiency grid as a function of the
smallest number of parameters that affect the recovery of a fake.
We can then use this multi-dimensional grid together with Monte Carlo
simulations to calculate the recovery efficiency of real transient events.

\subsection{Single parameter recovery efficiencies}
\label{sec:sing_rec_frac}

\begin{figure*}[p]
\epsscale{1.0}
\centering
\includegraphics[width=0.8\textwidth]{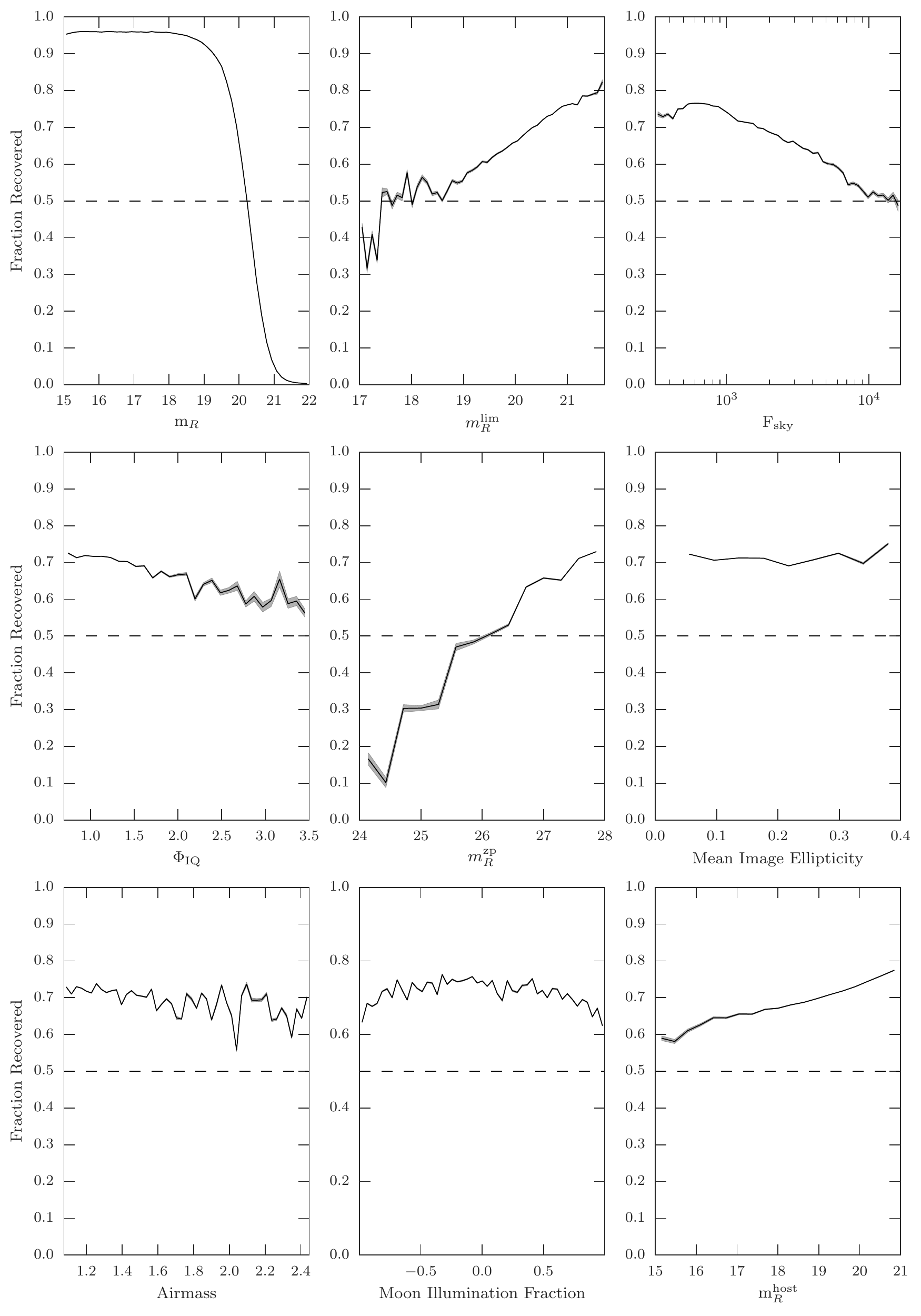}
\caption{The fake recovery efficiency $\epsilon$ as a function of 8 of the
  variables we consider and the fake's magnitude. In each individual panel, the recovery
  efficiencies are marginalized over the other parameters. These are
  (top row) the fake magnitude $m_R$, the limiting magnitude of the
  image \mlim, the median sky counts in the image \fsky, the
  ratio of the seeing in the science and reference images, the image
  photometric zeropoint \mzp, the mean image ellipticity,
  the airmass of the observation, the moon illumination fraction at
  the observing epoch and the host galaxy magnitude for the fake $m_R^\mathrm{host}$. The dashed lines represent the points at which 50\% of the fakes are recovered and the shaded regions contain 68.3\% of the probability.}
\label{fig:rec_all}
\end{figure*}

We begin by binning the data based on the input fake properties and
observing conditions with bin widths and number driven by the
precision with which the data are measured. For example, the \mzp\
values are determined by the real-time pipeline to an accuracy of
0.1\,mag, and so fewer, larger, bins are required compared to \mlim,
which is measured to a higher precision. The same binning is applied
to the equivalent data associated with the fakes that are recovered by
the PTF pipeline. We then define, in each bin $i$, the recovery
efficiency $\epsilon_i$ to be the ratio of the number of fake objects
recovered in each bin ($k_i$), to the total number of fakes originally
created in that bin ($n_i$) i.e., $\epsilon_i=k_i/n_i$.
One-dimensional recovery efficiencies for each variable are shown in
Figure~\ref{fig:rec_all}, in each case marginalized over the other
variables.

An important question is the calculation of uncertainties for each $\epsilon_i$. In each bin, the number of successful detections of a fake is a binomially distributed variable, i.e., there are $k$ successes (detections) out of $n$ independent trials (fakes), which is expressed by
\begin{equation}
\label{eqn:k_bino}
p(k|\epsilon,n)=\frac{n!}{k!(n-k)!}\epsilon^k(1-\epsilon)^{n-k}
\end{equation}
where the probability of success on each trial is the efficiency $\epsilon$. For the frequentist approach, it can be straight-forwardly shown that $\sigma_\epsilon=\sqrt{k(n-k)/n^3}$ \citep{paterno04}; however this equation fails in the limiting cases of $k=0$ or $k=n$. Instead, we use Bayes Theorem with Equation \ref{eqn:k_bino} to derive the posterior probability distribution of $\epsilon$
\begin{equation}
p(\epsilon|k,n)=\frac{\Gamma(n+2)}{\Gamma(k+1)\Gamma(n-k+1)}\epsilon^k(1-\epsilon)^{n-k}
\end{equation}
with a uniform prior in $\epsilon$ that $0\leq\epsilon\leq1$, where $\Gamma$ is the Euler gamma function; see \citet{paterno04} for details. Uncertainties are then calculated by numerically finding the shortest interval containing 68.3\% of the probability.

Several clear (and expected) trends are apparent in Figure~\ref{fig:rec_all}; for example fake
objects are more difficult to recover when fainter. However, even when the fake is bright
($m_R<18.5$\,mag), we note that a consistent $\approx3$\% of objects are
not recovered, implying that some small fraction of objects are missed no
matter what the brightness. Fake objects are also more difficult to recover as
the IQ of the science image becomes poorer relative to that of the
reference image; as the limiting magnitude becomes brighter; and as
the photometric zeropoint becomes brighter (i.e., the data have more
attenuation, presumably from clouds). The recovery fraction is also a
strong function of median sky counts (a brighter sky makes the fake
harder to detect), a weak function of the moon illumination fraction
(objects are harder to recover with a bright moon), and a weak function of
airmass (objects are marginally more difficult to recover at high
airmass). There is no measurable trend with image ellipticity,
indicating the image subtraction works well across most PTF data.

\subsubsection{Host galaxy surface brightness}
\label{sec:host-galaxy-sb}

As the fakes were inserted (see Section \ref{sec:fake-sn-locations}),
we record the total integrated $R$-band apparent magnitude of any host
galaxy ($m_R^\mathrm{host}$) from the \textsc{sextractor} catalog, as
well as the local surface brightness at the position of the fake. We
denote this latter parameter `\flbox', defined as the
background-subtracted sum of the pixel counts at the fake position
over different configurations of pixels. We record this metric in box
sizes from 1$\times$1 to 11$\times$11 pixels, however our default for
all \flbox\ measures is to use the integrated counts in a 3$\times$3
box as this is close in size to the PSF of a typical fake. This metric
provides local environment information for an object's recovery
efficiency, i.e., the transient detection pipeline's ability to
discover sources against a bright background. The \flbox\ metric
is the only parameter we discuss that was not output from the from
the real-time pipeline during survey operations between 2009-2012. Thus
any study based on the results of our efficiencies, which explicitly
require the use of \flbox, will need to measure \flbox\ for their
transient objects so that they are directly comparable to our fake simulations. The real-time pipeline did measure a fixed aperture flux of 5 pixels in both the subtraction and the reference, referred to as the {\it flux-ratio} in \citet{2012PASP..124.1175B}. However, while useful for computing the real-bogus score, we found it insufficient for our needs as it was in general too large compared to the typical PSF.

\begin{figure*}
\centering
  \epsscale{1.0}
\includegraphics[width=\linewidth]{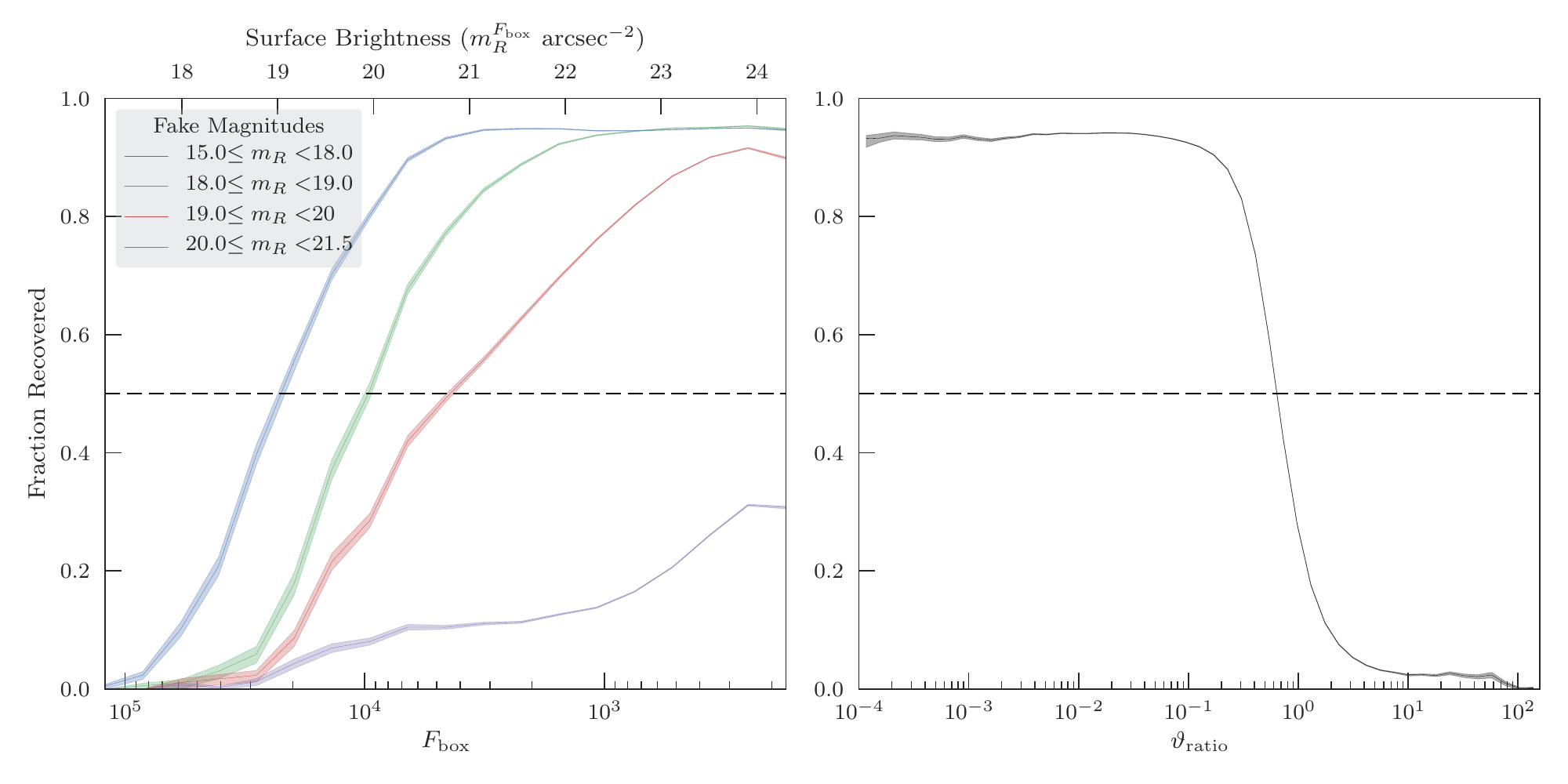}
\caption{Left: The recovery efficiency as a function of the \flbox\ parameter on the bottom axis and $m_{R}^{\flbox}$ surface brightness on the top axis. We see the expected trend of fewer fakes recovered when they are situated in bright regions (analogous to bright host galaxies). Right: The recovery efficiency as a function of the ratio of counts in \flbox\ to the counts of the fake point source (\frat). We find that if the local surface brightness is $\lesssim$0.1$\times$ than the fake, the pipeline finds it $\approx$95\% of the time. Beyond this the recovered fraction sharply falls off and the 50\% recovery fraction is at \frat$\approx$10$^{-0.2}$}
\label{fig:Fbox}
\end{figure*}

Figure~\ref{fig:rec_all} shows that fakes become more difficult to
recover in brighter galaxies. However, $m_R^\mathrm{host}$ is a poor
choice of metric shown only for information. It is not applicable to
all real transient events \citep[where the host association may be
uncertain; e.g.,][]{2006ApJ...648..868S,2016arXiv160406138G}, and can
be mis-leading if, say, a transient is well-separated from a bright
host galaxy. Instead, the information is more usefully encapsulated by
the \flbox\ metric. In Figure~\ref{fig:Fbox} (left) we inspect the
recovery efficiency as a function of \flbox\, split into bins of fake magnitude, and see the expected trend where fakes in regions of higher surface brightness are less likely to be recovered. We also extend this analysis to a new
parameter, \frat: the ratio \flbox\ to the flux from the fake. This
new parameter, when considered alone, provides an insight into how
cleanly the image subtraction has been performed, which can
particularly affect the fainter fakes on bright galaxies. We note that
\frat\ has a degeneracy with $m_R$ (as both include the counts from
the fake) and in Section~\ref{sec:mult-recov-effic} we do not use
$m_R$ in conjunction with \frat\ for this reason.

In Figure~\ref{fig:Fbox} (right) we examine the recovered fraction of
fakes as a function of \frat. We find the expected trend where fakes
that are located in an environment of high surface brightness relative
to the object itself are less likely to have been detected by the
pipeline. The pipeline maintains a consistently high ability to
discover the fakes whilst the fakes are $\approx$10$\times$\ brighter
than \flbox. The recovered fraction rapidly drops off after this
point, with 50\% recovered at \frat$\approx0.7$

\subsubsection{Efficiencies as a function of time}
\label{sec:efficiencies-versus-time}

\begin{figure}
\centering
  \epsscale{1.0}
\includegraphics[width=\linewidth]{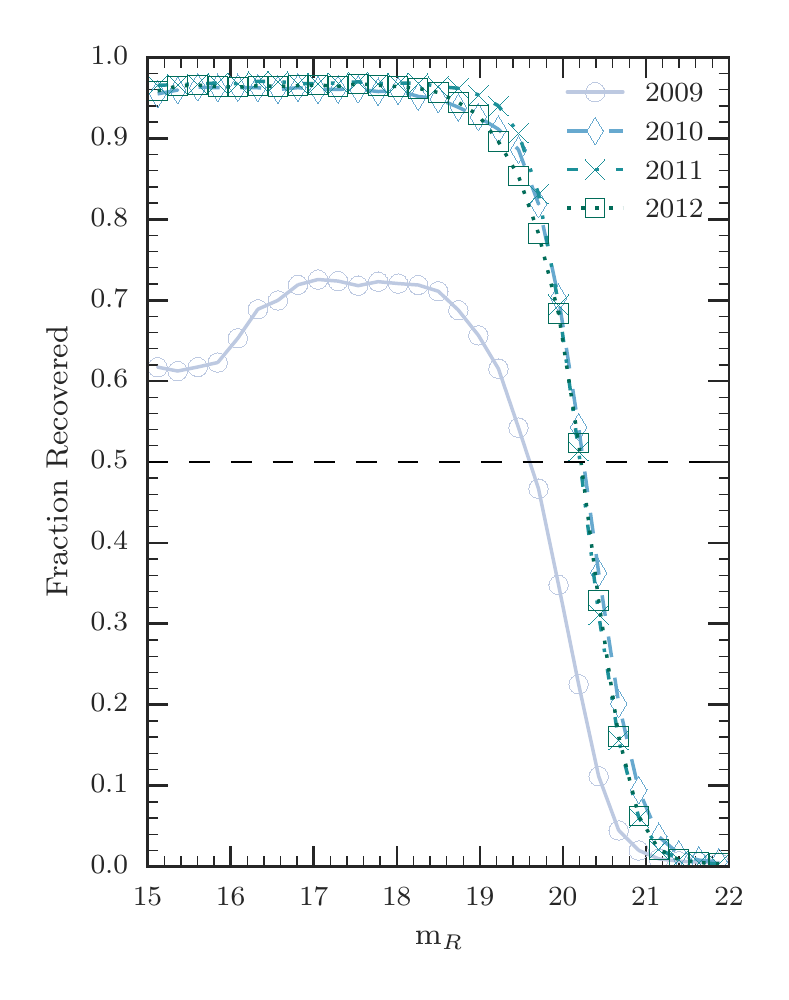}
\caption{The recovery efficiency as a function of fake apparent magnitude ($m_R$) for each year of the PTF survey (averaged over all observing conditions). The years 2010--2012 are consistent, but the year 2009 (the first year of the PTF survey) shows a large discrepancy (see discussion in section~\ref{sec:sing_rec_frac}). We exclude 2009 from our  analysis.}
\label{fig:yrsplit}
\end{figure}

Due to the improvement and updating of the reference images during the
survey (Section~\ref{sec:sub_pipe}), we expect the recovery
efficiencies to show a time dependence. We therefore plot the recovery
efficiencies as a function of $m_R$ for each year of the survey
(Figure~\ref{fig:yrsplit}), and find that 2009 has a significantly
lower recovery efficiency than the subsequent years. The later years
-- 2010, 2011, 2012 -- all show consistent trends.  Given the large
discrepancy between 2009 and the later years, we exclude 2009 from our study. While the effect in 2009 is partly explainable
due to the likely lower quality of the references during 2009 (both in
terms of depth and IQ), we also note that the data from 2009 suffered
from a `fogging' problem on the PTF camera window \citep[described in
detail in][]{2012PASP..124...62O}. This likely dramatically decreased
the efficiency of the survey in the parts of the image affected by the
fogging during that period.

\subsubsection{50\% recovery efficiencies}
\label{sec:50pc-recovery-efficiencies}

\begin{figure*}

\centering
\includegraphics[width=\linewidth]{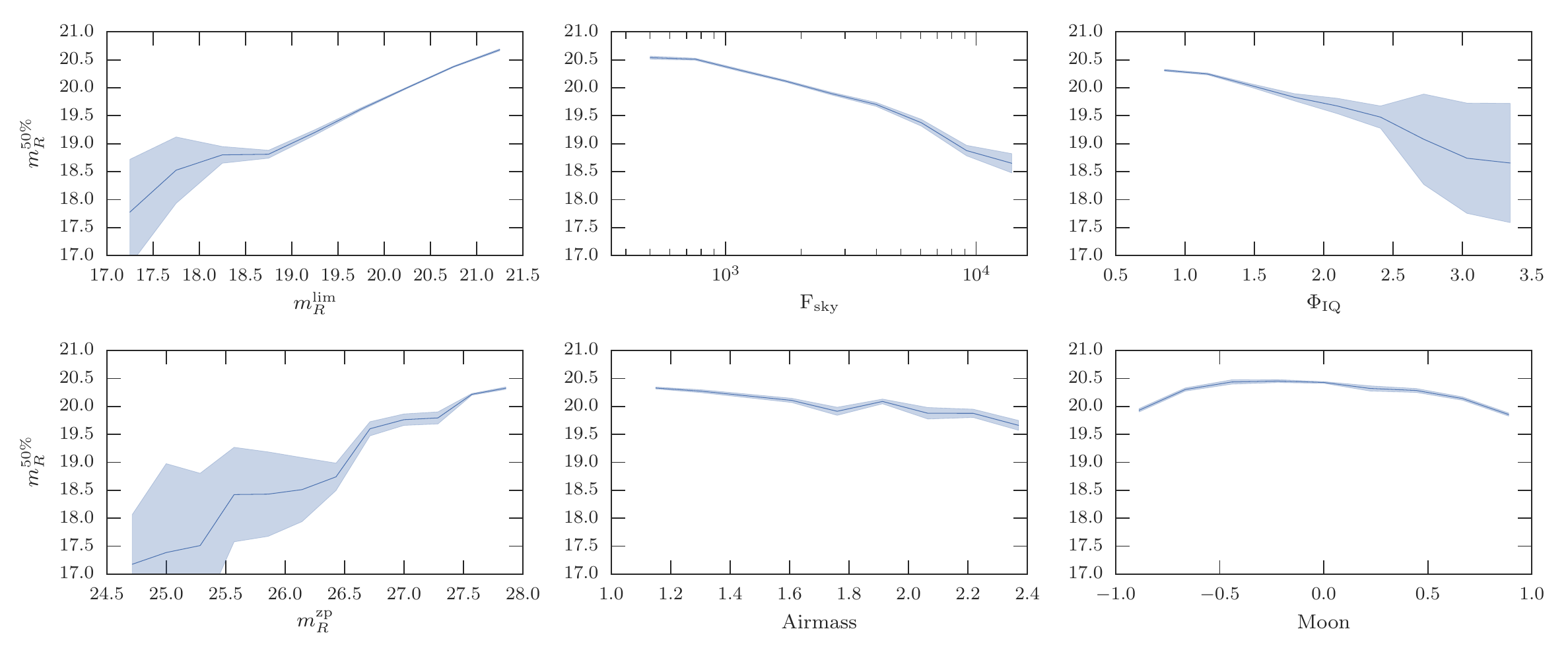}
\epsscale{1.0}
\caption{The 50\% recovery magnitude, $m_R^{50}$, as a function of
  various observing conditions. The results are plotted on the same
  axis scale to highlight the negligible dependence of these
  parameters in shifting the recovery efficiencies which is explained
  further in Section~\ref{sec:mult-recov-effic}. The shaded areas
  indicate the statistical uncertainties (containing
  68.3\% of the probability).}
\label{fig:mR50}
\end{figure*}

The 50\% recovery magnitude $m_R^{50}$ -- the magnitude at which PTF finds the same number of transients as it misses -- is another useful way of parameterizing the survey efficiencies. Taken over all observing conditions, $m_R^{50}\approx20.3$\,mag (Figure \ref{fig:rec_all}). However, $m_R^{50}$ depends strongly on the observing conditions and galaxy surface brightness. We show $m_{R}^{50}$ as a function of \mlim, \fsky, \phiiq\, \mzp, airmass and moon illumination fraction parameters in Figure~\ref{fig:mR50}; the trends are as expected.

\subsection{Multidimensional recovery efficiencies}
\label{sec:mult-recov-effic}

\begin{figure}
\centering
  \epsscale{1.0}
  \includegraphics[width=\linewidth]{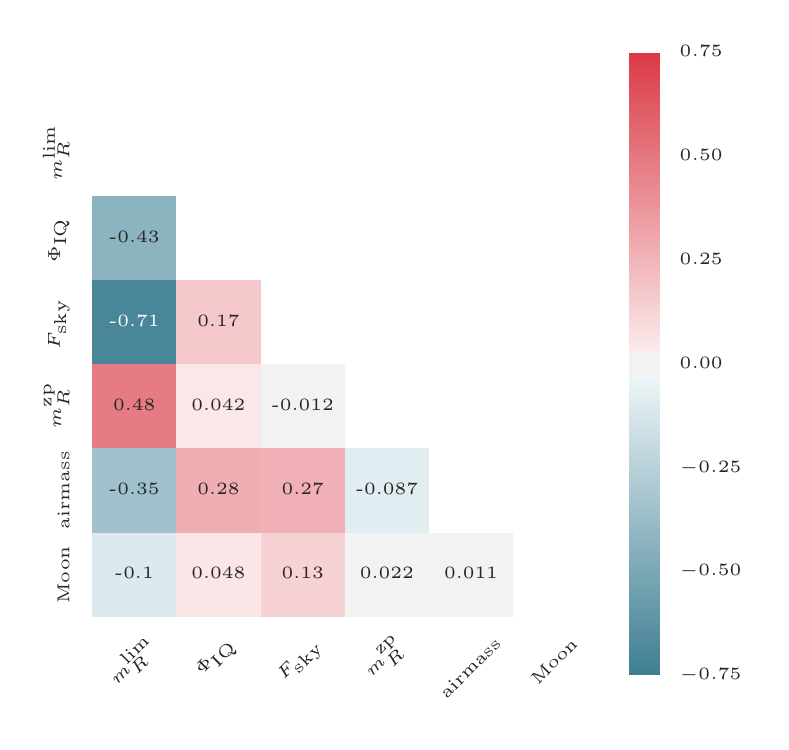}
  \caption{The correlation matrix for the observational metadata we
    record in our database. We explore the inter-dependencies of the
    parameters so that the number of dimensions in the final efficiency grid can be
    minimized to exclude strongly correlated parameters where no new
    information is gained. The values in each of the cells are the
    Pearson correlation coefficients.}
\label{fig:obs_cor_mat}
\end{figure}

We now extend our analysis of the single parameter recovery fractions
to study PTF's performance as a function of multiple variables -- our
final recovery efficiency grid. This method allows for situations to
be studied which cannot be encapsulated by any single parameter, for
example bright transients occurring in poor observing conditions. It
is possible to create a multi-dimensional efficiency grid from all of
the parameters discussed in Section~\ref{sec:sing_rec_frac} and shown
in Figure~\ref{fig:rec_all}; however, several of these variables are
likely to encapsulate similar information, and are therefore may be
degenerate (the correlations are given in Figure
\ref{fig:obs_cor_mat}). For computational reasons, it is more
efficient to construct a final recovery efficiency grid composed of
the fewest dimensions possible, but which capture the great majority
of the variation. In this section, we therefore examine the most
important variables that will make up a final efficiency grid. We
stress that whilst we aim to reduce the number of dimensions to
produce a final efficiency grid applicable for most purposes, there is
flexibility in this method to include any number of parameters to meet
the specific science goals of a study.

\begin{figure}
  \includegraphics[width=\linewidth]{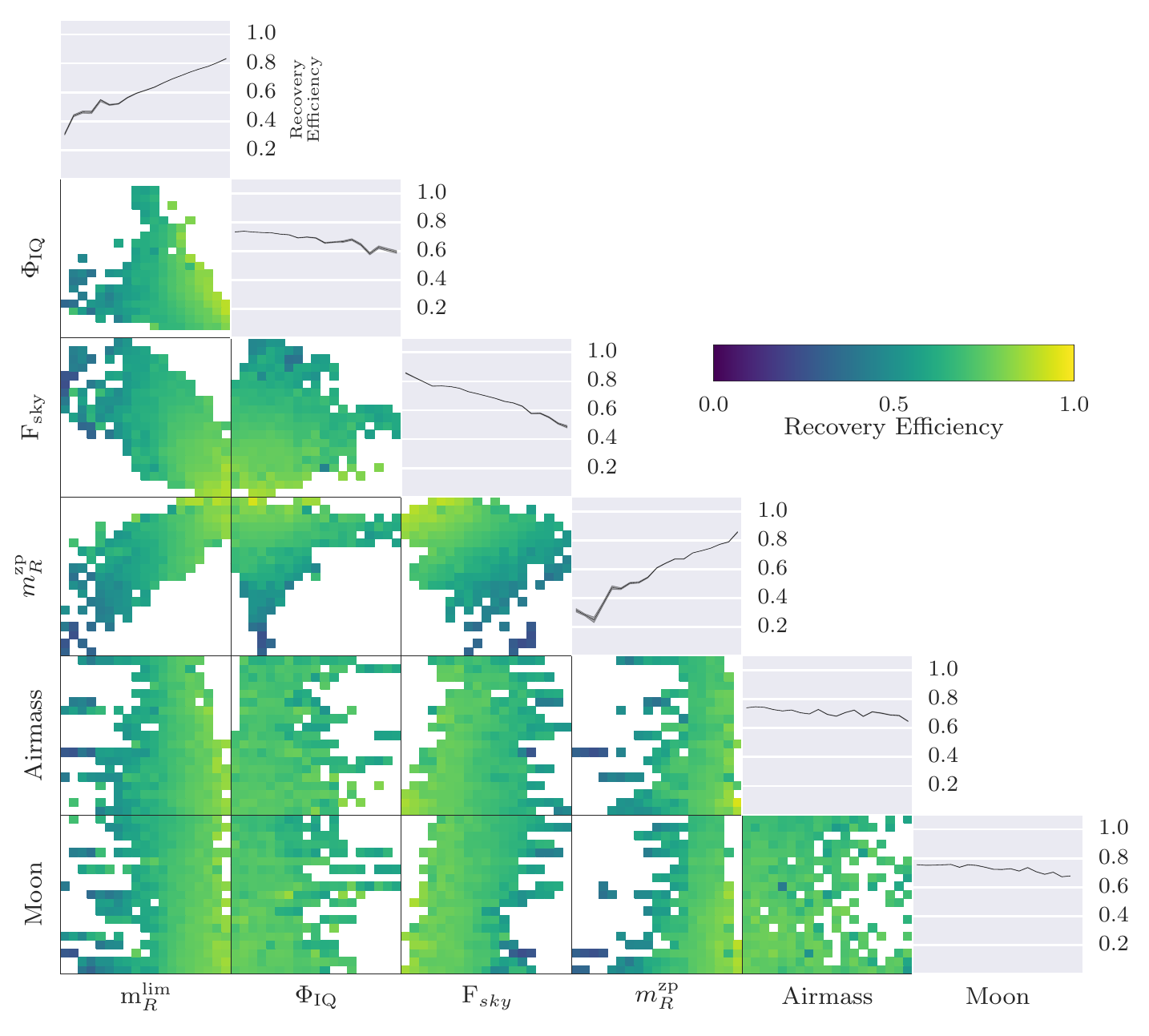}
  \caption{The 6-dimensional grid of the observing conditions metadata. The diagonal elements are the 1-dimensional recovery efficiencies $\epsilon$, projected along the axis of interest, with the gray shading denoting the area containing 68.3\% of the probability in $\epsilon$. The off-diagonal elements represent the different combinations of all the parameters.}
  \label{fig:norm_grid}
\end{figure}

The first dimension of our final efficiency grid is the apparent
magnitude of the fake object ($m_R$), a variable that is clearly
essential. the second dimension is \flbox, again containing information
not captured by the other variables. The remaining dimensions are then
drawn from the observing conditions. In Figure \ref{fig:obs_cor_mat},
we explore the Pearson correlation coefficients for the 6 pieces of
recorded metadata listed in Section \ref{sec:sub_pipe}; we neglect the
image ellipticity, as it has little impact on the efficiencies
(Figure~\ref{fig:rec_all}). We then construct, in Figure \ref{fig:norm_grid}, the 6-dimensional grid
of efficiencies where each cell in the grid is the probability of
recovering a transient as a combination of these 6 observing
conditions. These parameters are binned in an identical way to the
one-dimensional efficiencies as described in Section
\ref{sec:sing_rec_frac}, but with the absolute value of moon illumination fraction.

\begin{figure}
\epsscale{1.0}
\includegraphics[width=\linewidth]{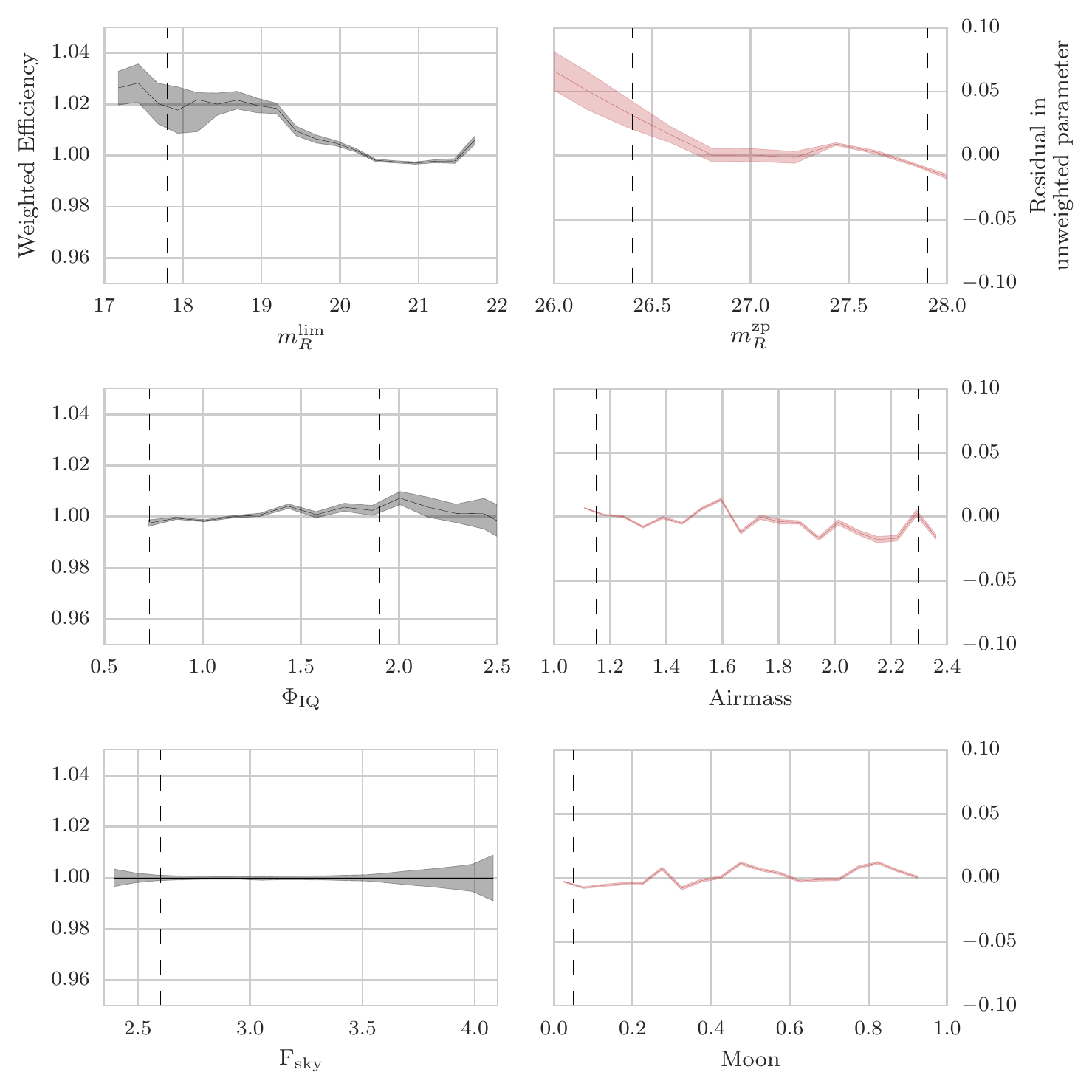}
  \caption{We explore which dimensions of the efficiency grid contain the most information. We show that by re-weighting the \mlim, \phiiq\ and \fsky\ axis (left column), we can remove the efficiency trends in the \mzp, airmass and $\left|\mathrm{moon\ illumination\ fraction}\right|$ axes. We show this by plotting the residuals from a perfect recovery efficiency (right column). The dashed vertical lines represent bounds containing 99\% of the data.}
  \label{fig:weight_residuals}
\end{figure}

To find the remaining dimensions with the most power, we weight each
multi-dimensional element in the grid by the inverse of the
1-dimensional detection efficiency associated with that bin for the
parameter we are interested in. We then assess the remaining
1-dimensional projections for indications of residual trends in
efficiency that would indicate that there is information in that axis
that was not also contained in the parameter used for the weighting.
We extend this analysis to combinations of weighted dimensions, and,
after experimentation, find from Figure \ref{fig:weight_residuals}
that we remove residual efficiency trends with \mzp, airmass and moon
illumination fraction when re-weighting the efficiency grid using the
\mlim, \phiiq, and \fsky\ parameters. (Note some residual trends
remain with \mzp, but only at the extremes of the distribution
representing poor observing conditions, presumably cloudy). 

\begin{figure}
\includegraphics[width=\linewidth]{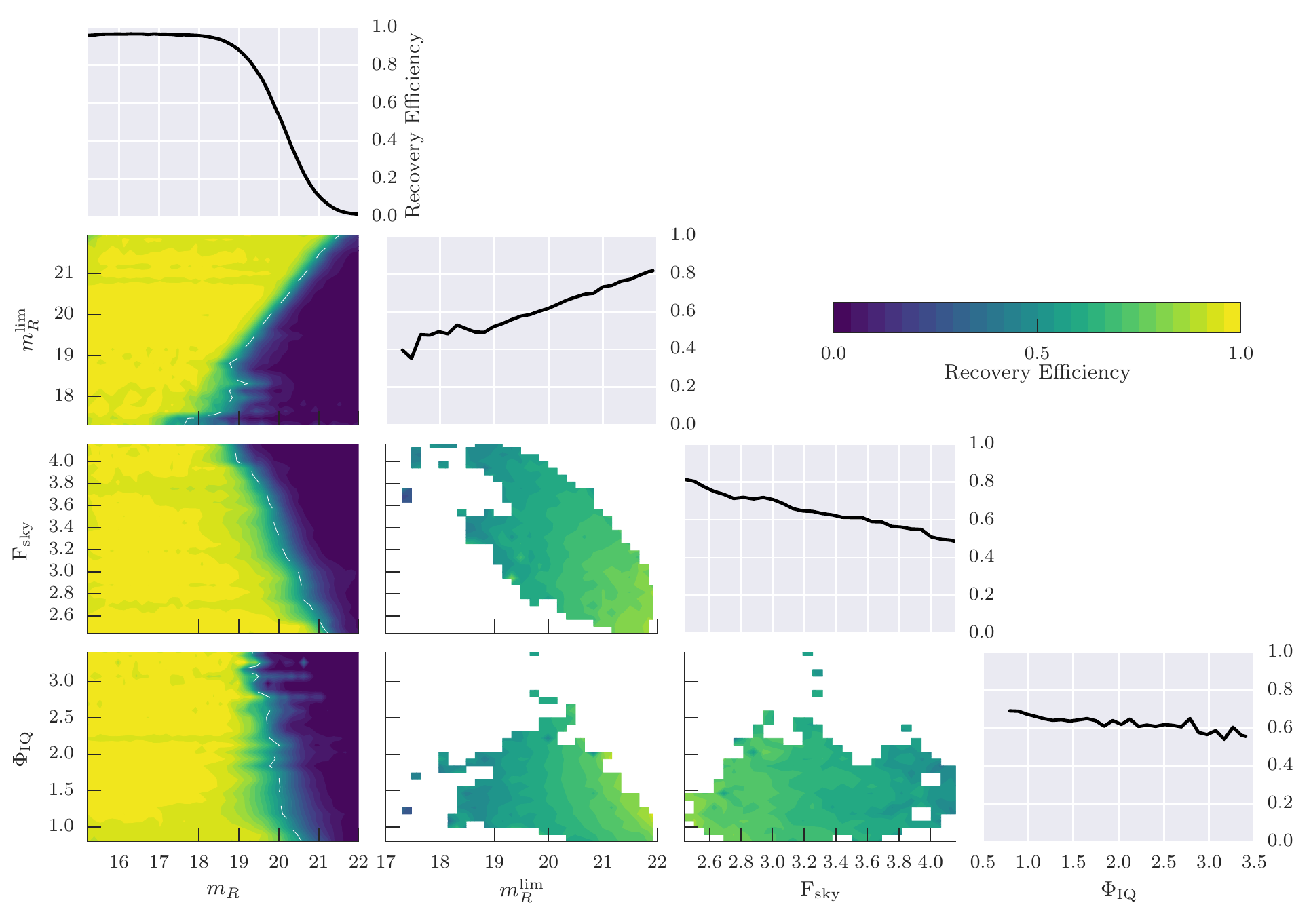}
\caption{The final multi-dimensional efficiency grid. The off diagonal entries show the two dimensional efficiencies for combinations of the parameters. The diagonal entries show the one dimensional recovery efficiencies created by marginalizing other the other grid parameters. The white dashed lines on the $m_{R}$ axis denote the 50\% recovery efficiency for this parameter against the other observing condition parameters.}
\label{fig:4D_Grid}
\end{figure}

Thus, the bulk of the variation in efficiency is captured by the 5
parameters of $m_R$, \flbox, \mlim, \phiiq, and \fsky, and the final
recovery efficiency grid is comprised of these variables (Figure
\ref{fig:4D_Grid}).  The reduced dimensionality of this final grid
also allows a finer binning of the data, increasing the resolution.
The grid can then be used to estimate the recovery efficiency of a
point source observed under any PTF observing conditions. This
probability of a detection, given $m_R$, \flbox, \mlim, \fsky, and
\phiiq, is calculated using a linear interpolation on the final
efficiency grid.

\section{Simulating PTF for a transient population}
\label{sec:survey_ops}

We have constructed a multi-dimensional recovery efficiency grid for
the PTF survey for transient point sources, describing the recovery
efficiency as a function of various astrophysical and observational
parameters. This allows us to calculate the fraction of point sources
recovered on any epoch or image from PTF as a function of the point
source magnitude $m_R$ and the host galaxy background. In this
section, we briefly describe how such an efficiency grid can be
applied to a real astrophysical problem; for example for calculating
the rates of particular types of transient events. We do this, in
effect, by simulating an artificial `night sky' across the PTF survey
area populated by transients defined by a time-dependent luminosity
model, and then exactly replicate PTF's observing pattern to observe
this artificial sky. Using the PTF metadata for each observation and
the efficiency grid from Section~\ref{sec:mult-recov-effic}, we can
then determine which of these simulated transients would have been
recovered.

Over the course of PTF, thousands of fields were observed across an
approximate footprint of $8000$\,deg$^2$. We initially explored
treating each PTF field as its own distinct area in which to simulate
transients. However, we found that this would underestimate our
calculation of the transient discovery efficiency as the PTF fields
spatially overlap, by design, and dither very slightly due to
imperfect telescope pointing. A transient event occurring in one of
the overlap regions would then be sampled more frequently under real
conditions than in the simulations, increasing the likelihood of
discovery and light curve coverage.

It is therefore simpler to treat the entire PTF survey as one single field, simulating transients at random positions within this field and with random explosion epochs. We use the PTF database to determine on which CCD (if any) the object would have been observed and the observing conditions for that CCD. These, along with the transient apparent magnitude, are used to interpolate on the multi-dimensional efficiency grid from Section~\ref{sec:mult-recov-effic}, to give the likelihood of recovering the transient on that epoch.

To determine whether a transient is observed on a given CCD, we use
the geospatial table extender PostGIS\footnote{http://postgis.net/}.
Each CCD is projected onto a spherical surface based on the RA and
Dec. of the corner pixels, and the geospatial location information is
stored and indexed in a new table along with the other PTF
observational metrics. The RA and Dec. of the simulated transient are
the query arguments which returns all CCDs which enclose that point.
With over 1.6$\times10^{6}$ observations taken throughout the survey,
this method allows us to retrieve all the CCDs, together with their
observing conditions, for a specific RA, Dec., and JD range, within
$\backsimeq0.01$\,s.

\subsection{Simulating a transient population}
\label{sec:sim_a_pop}

A transient population can be constructed from a time-dependent luminosity model and inserted into our artificial sky, and the efficiency grid then used to derive the probability that PTF would have discovered it. In this section we demonstrate this technique on a particular type of transient, type Ia supernovae (SNe Ia), a supernova class with a well-defined light curve model. Note that here we are simply demonstrating how the efficiency grid may be used; we apply our efficiency grid to a real SN Ia rates calculation in a later article.

The key to the method is to build up a second efficiency grid with its
axes made up of variables that describe the transient being simulated,
and that can be measured for real events. For this demonstration, we
use the SALT2 SN Ia model \citep{2007A&A...466...11G} within the
Python package \textsc{sncosmo} \citep{barbary_2014_11938} to generate
the SN Ia light curves. Our algorithm allows us to Monte Carlo
variations of the model and place them within the PTF survey at
different epochs and locations on the night sky. The model generates a
spectral energy distribution (SED) time series for a SN Ia, converted
into flux- or magnitude-space by integrating the SED through the
filter response of the $R_\mathrm{P48}$ filter.

For this demonstration, the key parameters are the light curve shape
\citep[the $x_1$ parameter, analogous to a light curve `stretch'; see
][]{1997ApJ...483..565P,2007A&A...466...11G} and the color ($c$, which
represents the \bv\ color of the SN at the time of maximum light).
Each simulated event also requires a spatial position and epoch of
explosion. To calculate the absolute magnitude of each event, $M_B$,
we randomly generate parameters from each SN Ia ($x_1$, $c$, $z$,
$\sigma_\mathrm{int}$) according to the distributions in Table
\ref{tab:sne_table} and insert them into
\begin{equation}
  \label{eqn:SALTeqn}
M_{B}=-19.05- \alpha x_1 + \beta C + \sigma_\mathrm{int}.
\end{equation}
where $\alpha$ and $\beta$ are `nuisance parameters' defining the $x_1$--luminosity and color--luminosity relations, $-19.05$ is the absolute magnitude for a typical SN Ia, and $\sigma_\mathrm{int}$ is the intrinsic dispersion of each event, capturing the intrinsic brightness variation in the SN Ia population after light curve shape and color correction. We use $\alpha= 0.141$ and $\beta= 3.101$ for this demonstration, following \citet{Betoule2014}. We then use the redshift $z$ to calculate the distance modulus to transform to an apparent magnitude in the observed $R_\mathrm{P48}$ filter, including Milky Way extinction according to the chosen spatial location on the sky. 

\begin{deluxetable}{lcc} 
\tablecaption{SALT2 SN Ia model parameters used in the simulation.\label{tab:sne_table}}
\tablecolumns{3}
\tablehead{   % column headings
  \colhead{Parameter} &
  \colhead{Distribution} &
  \colhead{Range}
}
\startdata
$x_1$ & Uniform & -3.0 to 3.0 \\
Color ($c$) & Uniform & -0.3 to 0.3 \\
Intrinsic dispersion ($\sigma_\mathrm{int}$)& Normal & $\mu=0, \sigma=0.15$ \\
Redshift ($z$)& Uniform & 0.0 to 0.1 \\
\enddata
\end{deluxetable}

\begin{figure*}[tb]
\epsscale{1.0}
\includegraphics[width=\textwidth]{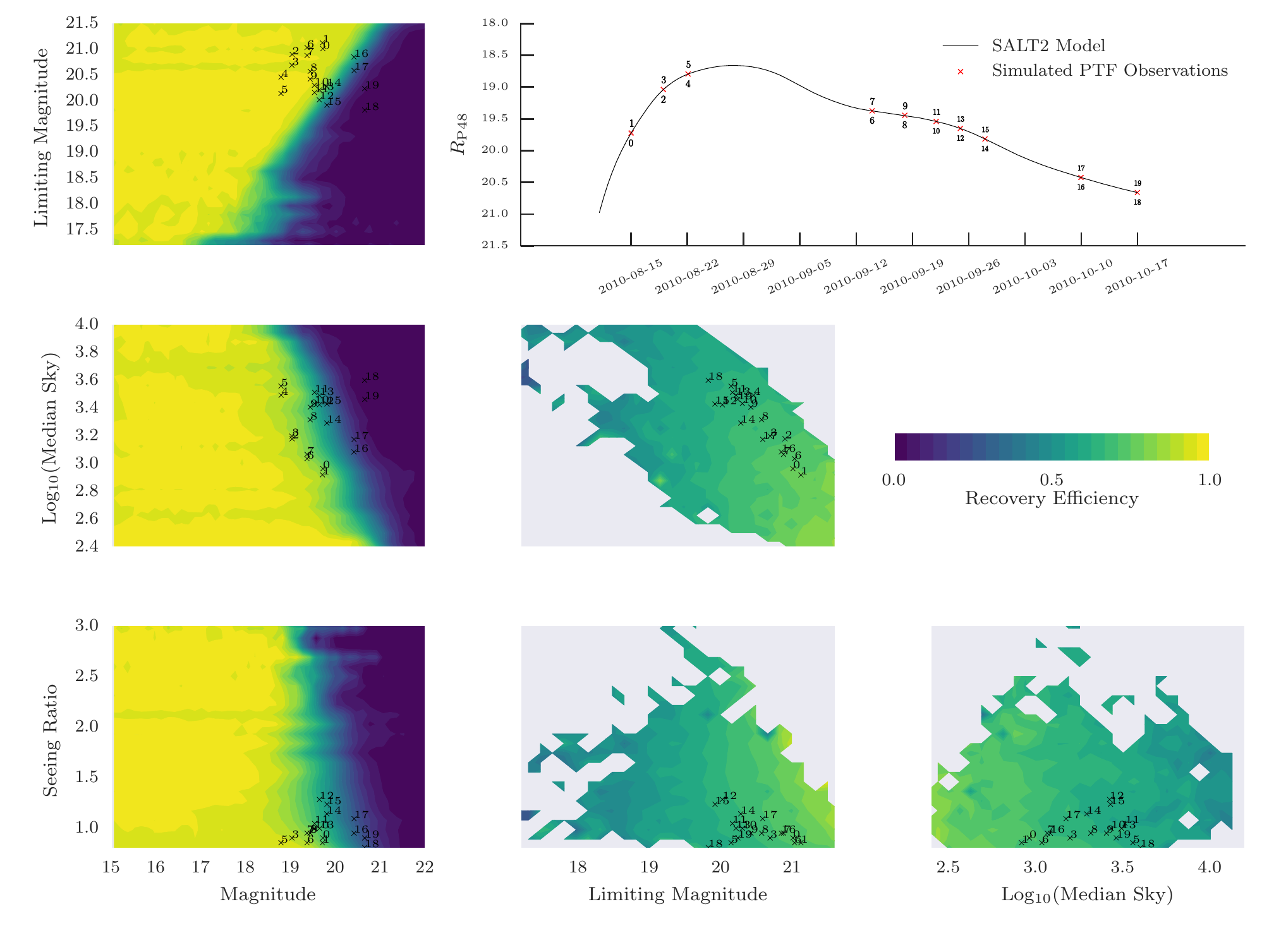}
\caption{A demonstration of the implementation of our detection efficiency grid for an example transient. The light curve model (in this case for a SN Ia) is used to predict the apparent magnitude on every epoch on which PTF made an observation of the spatial position of the model event (top right panel) and the PTF database returns the observing conditions on that epoch. Each combination of apparent magnitude and observing conditions then has an associated efficiency $P_\mathrm{detect}$, interpolated from the multi-dimensional efficiency grid; the figure shows the position of the point in various combinations of the grid dimensions, and the points are numerically labeled in the figure (PTF typically observes each position twice during a given night). If the value of $P_\mathrm{detect}$ is greater than or equal to a random number between 0 and 1, then that point is considered detected by PTF; otherwise the point is not considered detected. This process is repeated for each observation. The entire light curve can then be considered against appropriate selection criteria that determine the probability of whether the transient would be detected over the course of its evolution.}
\label{fig:LC_Eff_Grid}
\end{figure*}

This model then provides a light curve at a specific RA and Dec. on
our artificial night sky. A spatial query of the PTF database 
returns all the observing metrics for any CCD that could have observed
the SN, and the SN model gives the apparent magnitude for each of
these observing epochs. This observed magnitude is then used together
with the observing metadata to perform a multidimensional linear
interpolation on the efficiency grid described in Section
\ref{sec:det_eff}, returning the probability of PTF detecting the
object on each observed epoch ($P_\mathrm{detect}$). For each epoch,
we then randomly select a number, $\lambda$, from a uniform
distribution between 0 and 1 for comparison with that epoch's
detection probability: if $\lambda\le P_\mathrm{detect}$, the SN is
considered detected on that epoch, and if $\lambda>P_\mathrm{detect}$
the SN is considered not detected. Figure \ref{fig:LC_Eff_Grid}
demonstrates this concept, showing typical observational metric
locations on the efficiency grid for a demonstration SN Ia.

To construct recovery efficiencies as a function of the SN parameters,
we then construct a grid with the simulated SN parameters as the axes
of the grid (in this case $z$, $x_1$, $c$, $\sigma_\mathrm{int}$). By
simulating millions of fake SNe in the PTF area, simulated with
parameters drawn from the distributions in Table~\ref{tab:sne_table}
and assessing whether each would have been recovered by PTF, we can
then populate this grid. The recovery efficiency
$\epsilon_\mathrm{SN}$ of a real SN can then be estimated by
interpolating on this grid at the position of the values that
represent the real SN. For example, if a real supernova is found to
have $\epsilon_\mathrm{SN}=0.2$ from the simulated sky area, then it
means that this one object represents a population of five, where the
other four were missed by the survey.

Our method of simulating transients on an artificial sky and then
`observing' it encodes two pieces of information into the
$\epsilon_\mathrm{SN}$ metric. The first is an efficiency that is
intrinsically linked to the supernova model parameters. The second is
the sky area of the simulation; that is, the $\epsilon_\mathrm{SN}$
are calculated for an area of sky that may be larger than the area
actually observed, which must be borne in mind when interpreting the
$\epsilon_\mathrm{SN}$ values.
 
\section{Summary}
\label{sec:summary}

This paper has presented the transient detection efficiencies for the
Palomar Transient Factory (PTF). These efficiencies were quantified
through the addition of fake events into real PTF images, which were
then run through PTF's real-time transient detection pipeline. The
fraction of these fake transients recovered by the PTF pipeline then
quantifies the performance of PTF across a variety of observing
conditions, and transient magnitudes and local environments. This
information is captured in the form of a multi-dimensional efficiency
grid, which can then be used, together with Monte Carlo simulations of
transient events, to calculate rates and luminosity functions of
different transient types. We will detail these studies in later articles.

\acknowledgments

MS acknowledges support from EU/FP7-ERC grant no [615929] and STFC. CF
acknowledges the use of the IRIDIS High Performance Computing
Facility, and associated support services at the University of
Southampton, in the completion of this work. PEN acknowledges support
from the DOE under grant DE-AC02-05CH11231, Analytical Modeling for
Extreme-Scale Computing Environments.

Observations obtained with the Samuel Oschin Telescope and the 60-inch
Telescope at the Palomar Observatory as part of the Palomar Transient
Factory project, a scientific collaboration between the California
Institute of Technology, Columbia University, Las Cumbres Observatory,
the Lawrence Berkeley National Laboratory, the National Energy
Research Scientific Computing Center, the University of Oxford, and
the Weizmann Institute of Science.

This research used resources of the National Energy Research
Scientific Computing Center, a DOE Office of Science User Facility
supported by the Office of Science of the U.S. Department of Energy
under Contract No. DE-AC02-05CH11231.

We thank the anonymous referee for their useful comments.
\facility{PO:1.2m}

\bibliographystyle{aasjournal}

\end{document}